# Quantum Cost Efficient Reversible BCD Adder for Nanotechnology Based Systems

Md. Saiful Islam, *Member, IACSIT*, Mohd. Zulfiquar Hafiz, and Zerina Begum

*Abstract*—Reversible logic allows low power dissipating circuit design and founds its application in cryptography, digital signal processing, quantum and optical information processing. This paper presents a novel quantum cost efficient reversible BCD adder for nanotechnology based systems using PFAG gate. It has been demonstrated that the proposed design offers less hardware complexity and requires minimum number of garbage outputs than the existing counterparts. The remarkable property of the proposed designs is that its quantum realization is given in NMR technology.

*Index Terms*—Reversible Logic, Quantum Cost, Garbage Output, BCD Adder.

## I. Introduction

Irreversible logic circuits dissipate heat in the amount of $kT \ln 2$ Joule for every bit of information that is lost irrespective of their implementation technologies, where $k$ is the Boltzmann constant and $T$ is the operating temperature [1]. Information is lost when the input vector cannot be recovered from its corresponding output vector. Reversible logic circuit naturally takes care of heating because it implements only the functions that have one-to-one mapping between its input and output vectors. Therefore reversible logic design becomes one of the promising research directions in low power dissipating circuit design in the past few years and has found its application in low power CMOS design, digital signal processing and nanotechnology. According to [2] zero energy dissipation would be possible only if the network consists of reversible gates. Thus reversibility will become an essential property in future circuit design.

Reversible logic imposes many design constraints that need to be either ensured or optimized for implementing any particular Boolean functions [3-5]. Firstly, in reversible logic circuit the number of inputs must be equal to the number of outputs. Secondly, for each input pattern there must be a unique output pattern. Thirdly, each output will be used only once, that is, no fan out is allowed. Finally, the resulting circuit must be acyclic. Any reversible logic design should minimize the followings [6]:

- **Garbages**: outputs that are not used as primary outputs are termed as garbages.
- **Constants**: constants are the input lines that are either set to zero(0) or one (1) in the circuit's input side
- **Gate Count**: number of gates used to realize the system

- **Hardware Complexity**: refers to the number of basic gates (NOT, AND and EXOR gate) used to synthesize the given function
- **Quantum Costs**: quantum realization cost of the design in any particular nanotechnology

This paper presents a novel quantum cost efficient reversible logic implementation of BCD adder. The design includes PFAG as its basic building block proposed in [6]. The quantum realization cost of PFAG in NMR technology is 8 [6]. The proposed reversible BCD adder is optimized in terms of gate count and quantum costs.

## II. Background

### A. Reversible Gate

A gate or a circuit is called reversible if there is a one-to-one correspondence between its input and output assignments. Any reversible circuit realizes only the function that is reversible. There exist many reversible gates in the literature. Among them Feynman gate, FG [7], Peres gate, PG [8], Toffoli gate, TG [9], Fredkin gate, FRG [10] and Khan gate, NG [11] are mostly common (Fig. 1-5).

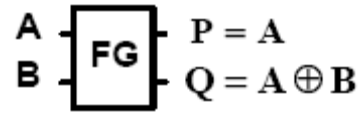

Fig. 1. Feynman Gate.

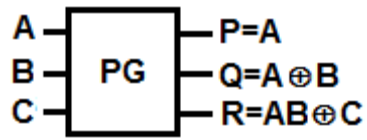

Fig. 2. Peres Gate.

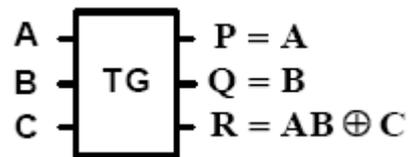

Fig. 3. Toffoli Gate.

The quantum realizations of all these gates are not available in the literature. Only FG, PG, TG, and FRG have been realized in nanotechnology. The detail cost of a reversible gate depends on any particular realization technology of quantum logic. Every permutation quantum gate is built from 1*1 (inverter) and 2*2 (FG) quantum primitives and its cost is calculated as a total sum of 2*2 gates. The quantum realization cost of FRG and TG is 5. The quantum cost of PG is 4 and it is the cheapest in terms of quantum realization cost [5].

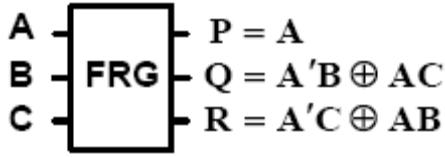

Fig. 4. Fredkin Gate.

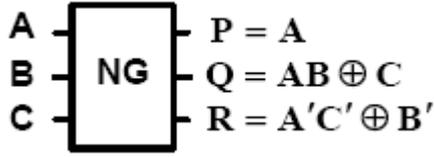

Fig. 5. Khan Gate.

*B. Reversible Full Adder Gate*

Full adder is the fundamental building block of many computational units. The anticipated paradigm shift logic compatible with optical and quantum requires compatible adder implementations. Minimization of reversible full-adder circuits and their implementation issues has been discussed in [3-5]. It has been shown that a full adder circuit can be realized with at least two garbage outputs and one constant-input [5].

It has been assumed that full adder circuit that can work singly as a reversible full-adder unit will be beneficial to the quantum realization of other complex systems. The family of reversible full adder gates are PFAG gate [6], TSG gate [12], MKG gate [13] and HNG gate [14] (Fig. 6-9). The quantum realizations of all these reversible full adder gate units are not available in the literature. Only PFAG has been realized in NMR nanotechnology. The quantum cost of PFAG is 8 [6]. It has also been demonstrated in [6] that PFAG is better than TSG, MKG and HNG in terms of hardware complexity.

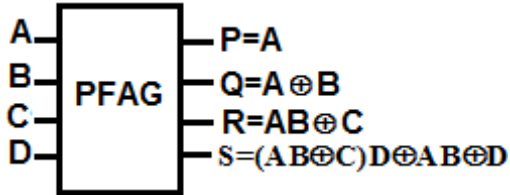

Fig. 6. PFAG Gate.

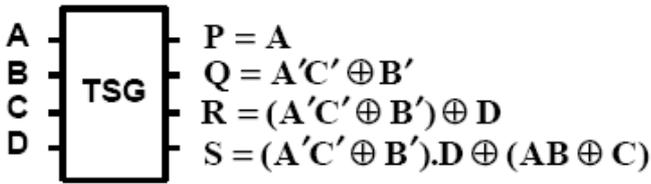

Fig. 7. TSG Gate.

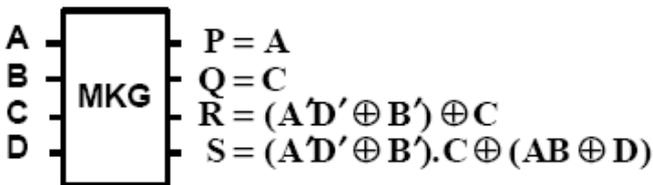

Fig. 8. MKG Gate.

### III. Reversible Logic Realization of BCD Adder

A BCD adder circuit adds two BCD numbers and converts the result into its equivalent BCD number. The conversion is needed because of the occurrence of overflow of the addition. To design a BCD adder we need 4-bit parallel adder. The reversible 4-bit parallel adder realized using PFAG gate shown in Fig. 10. It includes 4 PFAG and requires 4 constant inputs. The quantum realization cost of the proposed implementation is 32 and the design produces 8 garbage outputs.

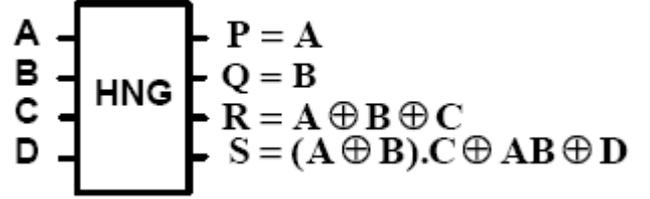

Fig. 9. HNG Gate

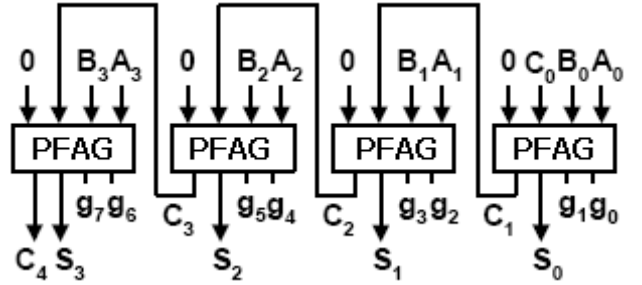

Fig. 10. Proposed Reversible 4-bits Parallel Adder

This study presents two BCD adder implementations realized using PFAG gates as its fundamental building block proposed in [6]. The designs are shown in Fig. 11 and Fig. 12. The proposed BCD adders have two reversible 4-bit parallel adders, which require total eight PFAG gates. Furthermore, two extra PFAG gates and one PG gate are required for implementing the correction logic. The first design also includes 4 FG gates to avoid fan-out of bits. To avoid fan out, the second design includes two FG gates and one HNFG gate proposed in [15]. The results are summarized in Table 1.

### IV. Result and Discussion

The proposed reversible BCD adder designs are efficient in terms of quantum cost than the existing counterparts. The quantum realizations of the existing BCD adders are not possible since those designs include gates that are not realized in nanotechnology. The following section demonstrates the superiority of the proposed designs in terms of hardware complexity, garbage outputs and constant inputs. The section will also discuss about quantum costs of the proposed designs.

*A. Hardware Complexity*

One of the main factors of a circuit is its hardware complexity. It can be proved that the proposed circuit is better than the existing approaches in terms of hardware complexity. Let

$\alpha$ = A two input EXOR gate calculation
$\beta$ = A two input AND gate calculation
$\delta$ = A NOT gate calculation

The total logical calculations of the proposed designs are T= $56\alpha+21\beta$. The total logical calculation of existing designs are T= $49\alpha+21\beta+6\delta$ [15], = $42\alpha+30\beta+33\delta$ [16], $59\alpha+30\beta+33\delta$ [17] and $75\alpha+48\beta+36\delta$ [17] (carry skip BCD adder) respectively. Therefore, we can say that the proposed design offers less hardware complexity than the existing counterparts.

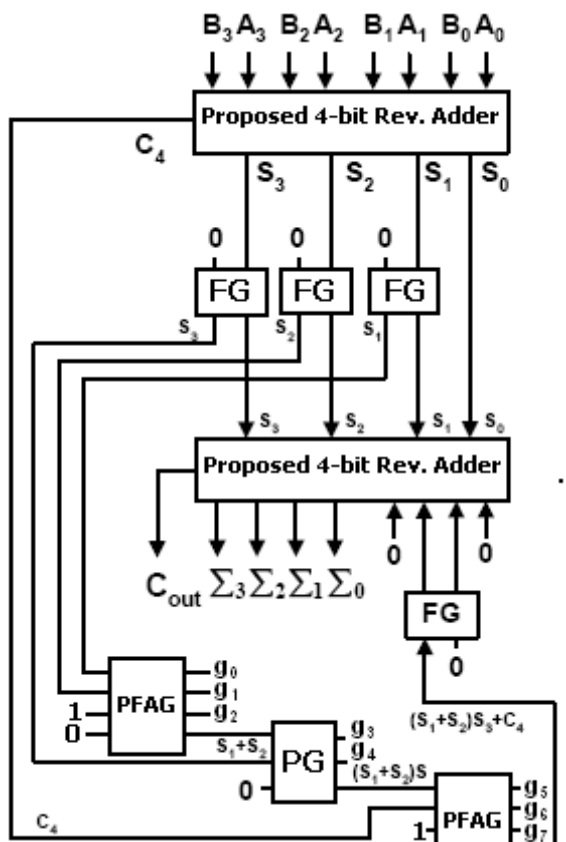

Fig. 11. Proposed reversible BCD adder: Design 1.

### B. Garbage Outputs

Garbage output refers to the output of the reversible gate that is not used as a primary output or as input to other gates [5]. One of the other major constraints in designing a reversible logic circuit is to lessen number of garbage outputs. The proposed design produces 24 garbage outputs that is less than the design proposed in [17]. But the proposed designs produce two extra garbage outputs than the designs presented in [15-16].

### C. Gate Count

The proposed design requires 14 reversible gates (Design 2) that are equal to the design given in [15]. The design presented in [16] requires 23 reversible gates and the design given in [17] requires 16 and 22 reversible gates (carry skip BCD adder) respectively.

### D. Constant Inputs

Number of constant inputs is one of the other main factors in designing a reversible logic circuit. The input that is added to an *nxk* function to make it reversible is called constant input [5]. The proposed designs require 19 constant inputs and the design given in [15] requires 17 constant inputs. This is because our designs include only those gates that can be realized in nanotechnology.

### E. Quantum Costs

The detailed cost of a reversible gate depends on any particular realization of quantum logic. The proposed designs include only those gates that have already been realized in nanotechnology namely FG, PG, HNFG and PFAG. The quantum realization cost of FG, PG and PFAG are 1, 4, and 8 respectively. HNFG basically incorporates the functionality of two FG gates and therefore its cost will be 2. The quantum realization costs of the designs presented in this study are 88. The existing counterparts of the presented reversible BCD adders have not been realized in any nanotechnology and therefore the costs of those designs are unknown at present.

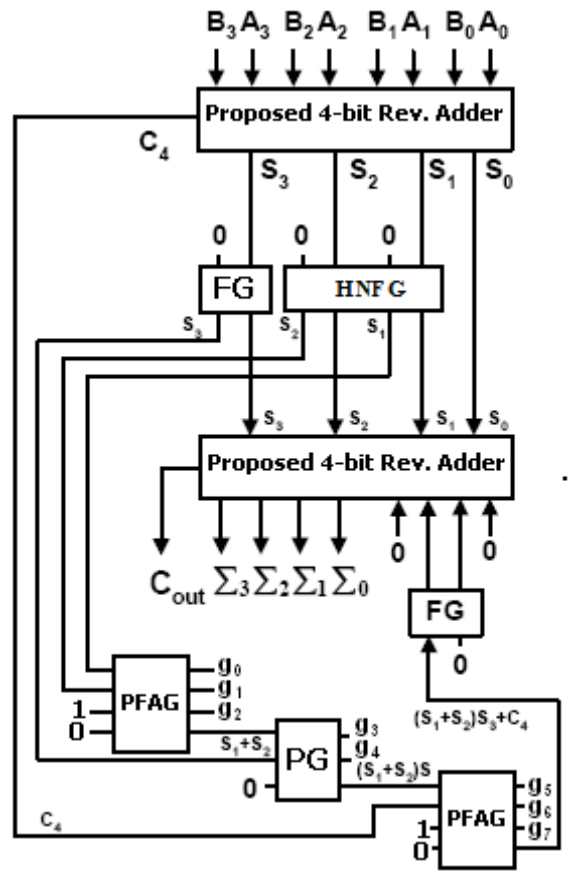

Fig. 12. Proposed reversible BCD adder: Design 2.

## V. CONCLUSION

This paper presents the efficient design of novel quantum cost efficient reversible logic implementation of two BCD adders that can be technologically mapped. The design includes only those reversible gates for which the quantum realization is known at present. The study also demonstrates its effectiveness in terms of hardware complexity, gate count, garbage outputs and constant inputs.

TABLE I: Comparative Analysis Of Different Reversible BCD Adders

| BCD Adder Design | Gate Count | No. of Garbage Outputs | Total Logical Calculations | Quantum Costs |
|---|---|---|---|---|
| This study: Design 1 | 10 PFAG +4FG+1PG=15 | 24 | 56 $\alpha+21\beta$ | 88 |
| This study: Design 2 | 10 PFAG+1PG +2FG+1HNFG=14 | 24 | 56 $\alpha+21\beta$ | 88 |
| BCD adder [15] | 8 HNG +2NG+ 1 TG+2FG + 1HNFG=14 | 22 | $49\alpha+21\beta+6\delta$ | Unknown |
| BCD adder [16] | 19+4FG=23 | 22 | $42\alpha+30\beta+33\delta$ | Unknown |
| Conventional BCD adder plus fanout [17] | 11+5FG=16 | 22 | $59\alpha+30\beta+33\delta$ | Unknown |
| Carry skip BCD adder plus fanout [17] | 15+7FG=22 | 27 | $75\alpha+48\beta+36\delta$ | Unknown |

**Md. Saiful Islam** is an Asst. Professor in the Institute of Information Technology at the University of Dhaka., Dhaka-1000, Bangladesh. He holds MS degree in Computer Science and Engineering from the University of Dhaka, 2007. Mr. Islam has more than 22 research articles published in several journals and conferences. Mr. Islam also leads and teaches modules at both BSc and MSc levels in Information Technology and Software Engineering. His research interests include Reversible Logic Design, Multimedia Information Retrieval, Machine Learning, Evolutionary Computing and Pattern Recognition.

**Mohd. Zulfiquar Hafiz** obtained his B. Sc (Hons) and M. Sc in Pure Mathematics from the University of Dhaka in 1986 and 1987 respectively.
Mr. Hafiz started his carrier as a programmer in the Computer Centre at University of Dhaka in 1993. He was repositioned as a Lecturer in the centre in 1997. He is still continuing to serve University of Dhaka as an Associate Professor in the Institute of Information Technology. His current research interest areas include Computer Networks, Embedded Systems and Computational Fluid Dynamics.

**Dr. Zerina Begum** obtained her B. Sc (Hons) and M. Sc in Physics from the University of Dhaka in1986 and 1987 respectively. Dr. Begum started her carrier as a Research Fellow in Bose Research Centre at University of Dhaka. Two years later she joined as a Programmer in the Computer Centre of the same university. She was awarded M. Phil for her research in Semiconductor Technology and Computerized Data Acquisition System in 1995. She was repositioned as a Lecturer in the centre in 1997. As recognition of her outstanding research in organic semiconductor, Department of Physics, University of Dhaka awarded her with Ph. D. degree in 2007. She is still continuing to serve University of Dhaka as a Professor in the Institute of Information Technology. Her current research interest areas include Semiconductor Technology, Computerized Data Acquisition System, Advanced Digital Electronics, Software Engineering and Computational Biology. Dr. Begum is looking forward to build up IIT as the vanguard of ICT education and research in South Asia.